\documentclass[twocolumn,showpacs,preprintnumbers,amsmath,amssymb]{revtex4}
\usepackage{amsmath,amsthm,amsfonts,graphicx,epsfig}
\usepackage[usenames]{color}

\newcommand{\dbar}{\kern-.1em{\raise.8ex\hbox{ -}}\kern-.6em{d}}

\def \be{\begin{equation}}
\def \ee{\end{equation}}

\begin{document}
\title{A Comment on ``Optimal Stroke Patterns for Purcell's Three-Link Swimmer"}
\author{  O. Raz  and J. E.  Avron,
\\
Department of Physics\\ Technion, 32000 Haifa, Israel}
\date{\today}%
%%\begin{abstract}
%%\end{abstract}
\maketitle

%%%%%%%%%%%%%%%%%%%%%%%%%%%%%%%%%%%%%%%%%%%%%%%%%%%%%%%%%%%%%%%%%%%%

In their letter ``Optimal strokes patterns for Purcell's
three-linked swimmer''  Tam and Hosoi \cite{Hosoi} describe strokes
that optimize  the distance  and swimming efficiency of Purcell's
three linked swimmer. The calculations are made in the framework of
Cox's slender body theory \cite{Cox} where the aspect ratio
$\kappa$, a.k.a. the slenderness,  is large.
\begin{figure}[h]
    \begin{center}
    \includegraphics[width=8cm]{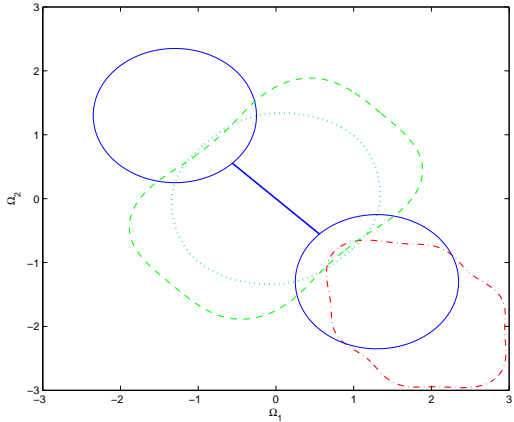}
    \end{center}
    \caption{ $\Omega_1,\Omega_2$  are the exterior angles made by the center
    link and the two arms. The approximate rectangle (green) represents
    the distance optimizer of moderate strokes of Tam and Hosoi.
    The (blue) dumbbell is a large stroke that outperforms the moderate stroke.
    The approximate (Light-blue) circle at center represents the efficiency
    optimizer of moderate strokes found by Tam and Hosoi.
    The off-center (red) wobble oval is a large stroke with higher efficiency.}
    \label{fig:OptStrokes}
\end{figure}

Tam and Hosoi find that the approximate square and circle at the
center of the figure are optimizers of the distance and efficiency
respectively. Here we want to point out that these optimizers,
with moderate strokes, are likely to be global optimizers only for
$\kappa$ that are moderately large. However,  when $\kappa$ is
sufficiently large, there are large strokes that outperform the
moderate optimizers. For example, the dumbbell in the figure (with
a narrow "corridor" of about 0.01 rad.) will cover about 1.14 the
distance covered by the approximate square, (in the {\em opposite}
directions). Though large it is never close to self intersection
(in contrast with the large stroke optimizer discussed in
\cite{JFM-Stone}).  Cox's slender body theory to first order is
valid when the interaction forces between the links are small
compared with the forces due to the motion of the links. This
imposes a constraint on the angles: $(\pi-|\Omega_j|)\ln\kappa\gg
1$. In the dumbbell case where $\pi-|\Omega|\geq\frac{\pi}{4}$,
the constraint  is $\kappa\gg 4$. If one does not restrict
$\Omega_j$ one finds that the optimizer hits the boundary,
$\Omega=\pi$, and so is outside of Cox's theory for any finite
$\kappa$.

The case of optimally efficient stroke is similar. Restricting
$\pi-|\Omega|\geq0.14$ (in this case Cox's slender body theory
imposes $\kappa\gg 1250$), we have calculated the ex-center wobbly
oval in the figure to be 1.53 times more efficient then the
centered approximate circle of Tam and Hosoi. (Here too the
optimal stroke hits the boundary $\Omega=\pi$).

It is an interesting open question to determine the minimal values
of $ \kappa$ where the large strokes of the type described here,
outperform the optimizers for moderate strokes describe in
\cite{Hosoi}.

%%%\begin{figure}
%%%    \begin{center}
%%%    \includegraphics[width=8cm]{OptEff}
%%%    \end{center}
%%%    \caption{The green dashed closed curve at the center is the most efficient symmetric
%%%    stroke found by algorithm of Tam and Hosoi. The blue closed curve off-center
%%%    breaks the reflection symmetry but has larger efficiency}\label{fig:OptEff}
%%%\end{figure}
%%It is a known fact that an optimizer can break a symmetry. If that
%%is the case then the optimizer is not unique. Indeed, the optimizer
%%shown in Fig.~\ref{fig:OptStrokes} is degenerate with the path
%%symmetric under reflection in the diagonal.

%%As a guide for choosing the initial anzatz we have relied on the
%%curvature \cite{Wilczek-Shapere}.

%%%%%%%%%%%%%%%%%%%%%%%%%%%%%%%%%%%%%%%%%%%%

%%%%%%%%%%%%%%%%%%%%%%%%%%%%%%%%%%%%%

{\bf Acknowledgment} We thank A. Lishanskey, O. Kenneth for
discussions; A. Hosoi and D. Tam for helpful correspondence and
the ISF and the Technion VPR fund for support.

%%%%%%%%%%%%%%%%%%%%%%%%%%%%%%%%%
%%%%%%%%%%%%%% bib %%%%%%%%%%%%%%
%%%%%%%%%%%%%%%%%%%%%%%%%%%%%%%%%
%\begin{thebibliography}
%%\bibliographystyle{plain}
%%\bibliography{bibtex_for_purcell}
%%%\end{thebibliography}

\end{document}